\documentclass[11pt,a4paper]{article}
\usepackage[utf8]{inputenc}
\usepackage{amsmath}
\usepackage{amsfonts}
\usepackage{authblk}
\usepackage{amssymb}
\usepackage{subcaption}
\usepackage{float}
\usepackage{graphicx}
\usepackage[left=1.500cm, right=1.500cm, top=2.00cm, bottom=2.00cm]{geometry}
\usepackage[colorlinks = true, linkcolor = blue, urlcolor  = blue, citecolor = blue, anchorcolor = blue]{hyperref}
\usepackage{listings}
\usepackage{cleveref}
\usepackage{xcolor}

\lstdefinestyle{chstyle}{%
backgroundcolor=\color{gray!12},
basicstyle=\ttfamily\small,
commentstyle=\color{green!60!black},
keywordstyle=\color{magenta},
stringstyle=\color{blue!50!red},
showstringspaces=false,
numbers=left,
numberstyle=\footnotesize\color{gray},
numbersep=10pt,
stepnumber=1,
tabsize=2,
frame=Ltb,
framerule=1pt,
rulecolor=\color{black},
breaklines=true
}
\author{Kiran Thengil  \footnote{email:kirantnair94@gmail.com}}
\affil{SISSA, Via Bonomea 265, I-34136 Trieste, Italy}
\date{}
\title{ Number-operator-based inverse engineering technique in a two level system}
\begin{document}
\maketitle
{\bf Abstract:} This paper experimentally realize a new method for shortcuts to adiabaticity, number operator based inverse engineering method (NOBIE), using quantum computers built with transmon qubits. Digitized control pulses are programmed in an open source software development kit, qiskit, to execute the shortcut protocols. The obtained results shows the robustness of the NOBIE method, even though it is tested for an effective Hamiltonian of a qubit irrespective of the interaction with other qubits and noise associated with the control pulses.  
\section{Introduction}
One of the key challenges in the field of quantum computing lies in achieving adiabaticity, which refers to the ability to maintain the system in its initial eigen state throughout a computational process~\cite{TameemAdiabatic2018}. Adiabatic quantum computation holds great promise for solving complex optimization problems, but it often suffers from lengthy computation times~\cite{DongQuantum2022,ZhiminRealization2023}. To overcome this limitation, researchers have actively explored methods of shortcuts to adiabaticity (STA) as an alternative approach. STA techniques aim to accelerate the adiabatic evolution by manipulating the system's Hamiltonian through external control parameters~\cite{YinShortcuts2022,TakahashiDynamical2022,TorronteguiShortcuts2013,GueryShortcuts2019,delFocus2019}.

A quantum adiabatic process is an ideal concept that is not experimentally feasible due to its long evolution time. Shortcuts to adiabaticity (STA) protocols drive the quantum system to a desired adiabatic state within a finite time duration. STA protocols are formulated using various methods, such as the counter-diabatic method~\cite{MustafaAdiabatic2003,MustafaAssisted2005,BerryTransitionless2009,AdolfoShortcuts2012}, local counter-diabatic method~\cite{IbezMultiple2012,IevaCounterdiabatic2023}, fast forward method~\cite{ShumpeiFast2008,ShumpeiFast2010}, and invariant-based inverse engineering method~\cite{LewisAn1969,ChenFast2010,Torrontegui2011,ChenLewis2011,ChenOptimal2011,TorronteguiFast2012}. In addition to these methods, we propose a new method, the number operator-based inverse engineering (NOBIE) method and verify it experimentally using IBM quantum computers. The protocol obtained from the NOBIE method drives the quantum system along an entirely adiabatic path. NOBIE method is easy to develop as it does not necessitate the exact knowledge of the system Hamiltonian's eigenstates to construct the driving protocols.

This paper aims to propose and verify the NOBIE method for a single qubit using a quantum computer built with transmon qubits. We employ an open-source software development kit called qiskit to program the qubit's control fields and execute the experiment remotely using the online IBM Quantum Lab~\cite{DavidQiskit2018,ThomasQiskit2020}. The objective of this work is to test the NOBIE protocols for an effective Hamiltonian of the qubit, enabling an understanding of the method's accuracy independent of external interactions and control pulse noise. Furthermore, the paper demonstrates the reliability of digitized pulses for shortcuts in state-of-the-art quantum computers.

Section 2 of the paper provides an explanation of the general formalism of the NOBIE method and develops shortcut protocols for a general two-level system. In Section 3, a detailed description of adiabatic, non-adiabatic, and NOBIE methods for a qubit controlled by electromagnetic pulses is presented. The experimental realization of the NOBIE method using IBM Quantum Lab is discussed in Section 4. Finally, Section 5 concludes the paper, summarizing the findings and implications of the study.
\section{Number-operator-based inverse engineering for a general two-level system}
\label{NOBIEFAGTLS}
\subsection*{Formalism of the NOBIE method}
This section provides the theoretical formulation for NOBIE method to drive the system through the entirely adiabatic path. Consider the spectral decomposition of a system Hamiltonian with its instantaneous eigenstates~\cite{SakuraiModern2017},
\begin{equation}
H(t)=\sum_{n}\lambda_{n}(t)\vert n(t)\rangle\langle n(t)\vert,
\label{DecomposedHamil}
\end{equation}
where $\lambda_{n}(t)$ represents the instantaneous eigenvalues and $\vert n (t)\rangle$ denotes the instantaneous eigenstates. If it is possible to write the instantaneous eigenvalues in a separable form as a product of a function of the quantum number and a function of time, i.e., $\lambda_{n}(t)=f(n)f(t)$, then by dividing \autoref{DecomposedHamil} with $f(t)$, we can define a number operator,
\begin{equation}
\mathcal{N}(t)=\frac{H(t)}{f(t)}=\sum_{n}f(n)\vert n(t)\rangle\langle n(t)\vert.
\label{NumberOperator}
\end{equation}
The eigenvalues of the above operator, $f(n)$, depends only on the quantum number of state, $\vert n(t)\rangle$ and it is the inspiration to term the operator as number operator. The instantaneous eigenstates, $\vert n(t)\rangle$ can be used to decompose the general solution of the Schrodinger equation, $\vert \Psi(t)\rangle$ corresponding to the system Hamiltonian as  $\vert \Psi(t)\rangle=\sum_{n}c_{n}(t)\vert n(t)\rangle$, where $c_{n}(t)$ represents the time-dependent complex amplitudes~\cite{SakuraiModern2017}. Then, the expectation value of the number operator at any instant of time will be
\begin{equation}
\langle \mathcal{N}(t)\rangle=\langle \Psi(t)\vert\mathcal{N}(t)\vert \Psi(t)\rangle=\sum_{n}p_{n}(t)f(n),
\label{ExpectNumberOpertor}
\end{equation}
where $p_{n}(t)=c^{2}_{n}(t)$ is the time-dependent probability of occupation among the instantaneous eigenstates of the Hamiltonian. $p_{n}(t)$ is also an indicator of the adiabaticity of the evolution path~\cite{DanielGeneral2009,KatoOn1950,BornBeweis1928}. For an entirely adiabatic path, the probability, $p_{n}(t)$ among the instantaneous eigenstates of the Hamiltonian remains the same throughout the evolution. According to \autoref{ExpectNumberOpertor}, the expectation value of the number operator becomes time-independent for constant $p_{n}(t)$. Thus, for distinct $f(n)$ corresponding to each value of $n$, the number operator, $\mathcal{N}(t)$ will be an invariant if the system is driven through the entirely adiabatic path.  In NOBIE method, the number operator act as an invariant, such that the system under the inverse engineered NOBIE Hamiltonian always sweeps through the adiabatic subspace. The procedure to find the NOBIE Hamiltonian, $H_{N}(t)$ is as follows.\\
\textbf{Step 1:} Find the number operator for the system under consideration using \autoref{NumberOperator}.\\
\textbf{Step 2:} Guess the expected form of $H_{N}(t)$ using arbitrary time-dependent parameters by considering its hermiticity and the commutation compatibility with the number operator.\\ 
\textbf{Step 3:} Verify the invariance condition with the number operator,
\begin{equation}
\frac{\partial \mathcal{N}(t)}{\partial t}-i\left[\mathcal{N}(t), H_{N}(t)\right]=0,
\label{NumberInvariance}
\end{equation}
which gives the explicit form of $H_{N}(t)$. This step results in several conditions connecting the arbitrarily guessed parameters of $H_{N}(t)$ and the control parameters of $H(t)$. 
\par
In the following subsection, we find the NOBIE Hamiltonian for general two level quantum systems.

\subsection*{NOBIE for general two level systems}
A combination of three Pauli spin operators, $\sigma_{x}, \sigma_{y}$, and $\sigma_{z}$, can constitute a general two-level (or spin $\frac{1}{2}$) system Hamiltonian~\cite{SakuraiModern2017},
\begin{equation}
H(t)=x\sigma_{x}+y\sigma_{y}+z\sigma_{z},
\label{HamOfGenQub}
\end{equation}
where $x,y$, and $z$ are the time-dependent control parameters of the Hamiltonian. The spectral decomposition of the above Hamiltonian is same as given in \autoref{DecomposedHamil} and the instantaneous eigenvalues are in a separable form, $\lambda_{n}=f(n)f(t)$, where
\begin{equation}
\begin{split}
f(n)&=\pm 1\\
f(t)&= \sqrt{x^{2}+y^{2}+z^{2}}.
\end{split}
\end{equation}
Using the above separable form of $\lambda_{n}$, the explicit formula for the number operator becomes
\begin{equation}
\mathcal{N}(t)=\frac{H(t)}{f(t)}=\frac{x\sigma_{x}+y\sigma_{y}+z\sigma_{z}}{\sqrt{x^{2}+y^{2}+z^{2}}}.
\label{NumOpeForTwoLev}
\end{equation}
The denominator in the above equation is a function of time alone. Therefore, $\mathcal{N}(t)$ always commutes with the system Hamiltonian, $H(t)$, such that both the operators share the same eigen space at any instance of evolution. This property of number operator forms the basis for complete adiabaticity throughout the evolution. 
\par
We found the number operator as per the developed theory for two level systems. Further, we need to guess an arbitrary form of NOBIE Hamiltonian, $H_{N}(t)$. The system Hamiltonian, $H(t)$ is constituted using the Pauli spin operators. All the three Pauli spin operators span the Lie algebra space and obey the commutation relation, $\left[\sigma_{j},\sigma_{k}\right]=2i\epsilon_{jkl}\sigma_{l}$, where $\epsilon_{jkl}$ is the structure constant. The closed Lie algebra space necessitates the combination of all three Pauli operators for $H_{N}$. Assuming an explicitly time-dependent NOBIE Hamiltonian, $H_{N}(t)=a\sigma_{x}+b\sigma_{y}+c\sigma_{z}$ ($a,b$ and $c$ are functions of time), we can verify the invariance of the number operator using \autoref{NumberInvariance} and obtain the following conditions,
\begin{equation}
\begin{split}
\frac{f(t)\dot{x}-\dot{f}(t)x}{f(t)}&=2\left(zb-yc\right)\\
\frac{f(t)\dot{y}-\dot{f}(t)y}{f(t)}&=2\left(xc-za\right)\\
\frac{f(t)\dot{z}-\dot{f}(t)z}{f(t)}&=2\left(ya-xb\right)\\.
\end{split}
\label{MainConditions}
\end{equation}
Satisfying the above mutually dependent relations, the NOBIE Hamiltonian can drive the system through the entirely adiabatic path. Explicit solutions for the arbitrary parameters $a$, $b$ and $c$ can be obtained from the above equations.
\par
There are two methods to obtain the explicit solutions for the parameters $a,b$, and $c$, namely, mutually dependent solution and mutually independent (the complete derivation of both the methods is provided in \nameref{Appendix A}). The mutually dependent solution for \autoref{MainConditions} gives the expressions for the three parameters of $H_{N}(t)$ as
\begin{equation}
\begin{split}
a&=\frac{1}{2f^{2}(t)}\left(2x\left(xa+yb+zc\right)+\left(y\dot{z}-z\dot{y}\right)\right)\\
b&=\frac{1}{2f^{2}(t)}\left(2y\left(xa+yb+zc\right)+\left(z\dot{x}-x\dot{z}\right)\right)\\
c&=\frac{1}{2f^{2}(t)}\left(2z\left(xa+yb+zc\right)+\left(x\dot{y}-y\dot{x}\right)\right).		
\end{split}			
\end{equation}	
Many arbitrary functions might be suitable for the above defined $a,b$ and $c$, which may result in distinct $H_{N}(t)$. The mutual dependence of the parameters makes it difficult to design the time-dependence of the parameters. However, assuming the value of $xa+yb+zc=f^{2}(t)$ at any instant of evolution gives the NOBIE Hamiltonian,
\begin{equation}
H_{N}(t)=H(t)+\frac{\vec{r}\times \frac{d\vec{r}}{dt}}{2f^{2}(t)}\cdot\vec{\sigma},
\label{CDHamiltonian}
\end{equation}
where vectors, $\vec{r}=(x,y,z)$ and $\vec{\sigma}=(\sigma_{x},\sigma_{y},\sigma_{z})$. This particular case of NOBIE Hamiltonian is equivalent to the counter-diabatic Hamiltonian for a general spin $\frac{1}{2}$ system~\cite{BerryTransitionless2009}.
\par
A slightly different approach provide us with all the three parameters of $H_{N}$ independent of each other. A set of such mutually independent solution is
\begin{equation}
\begin{split}
a&=0\\
b&=\frac{\dot{f}(t)z-f(t)\dot{z}}{2xf(t)}\\
c&=\frac{f(t)\dot{y}-\dot{f}(t)y}{2xf(t)},
\end{split}
\label{IndependentSolutions}	
\end{equation}
which reduced the NOBIE Hamiltonian to $\sigma_{y}$ and $\sigma_{z}$ components. It is possible to reduce the NOBIE Hamiltonian to any two components of Pauli operators using this procedure to achieve the shortcut to adiabaticity with reduced number of controls. The reduction of number of controls will be extremely useful for experimental implementation of shortcut to adiabaticity and it is an exclusive characteristic of NOBIE method.

\section{Shortcut to adiabaticity in a qubit controlled by electromagenetic pulses}
\subsection*{Adiabatic dynamics}
\begin{figure}[H]
\centering
\includegraphics[width=0.4\textwidth]{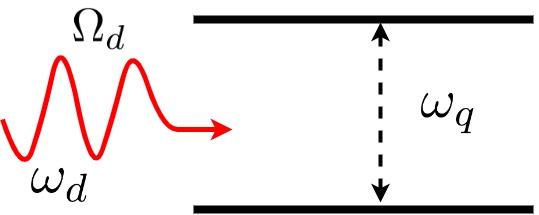}
\caption{Schematic diagram of qubit controlled by electromagnetic wave. The qubit frequency is $\omega_{c}$, the driving frequency of the pulse is $\omega_{d}$, and $\Omega_{d}$ is the time-dependent amplitude of the pulse.} 
\label{Qubitpulse}
\end{figure}

\begin{figure}[H]
\centering
\includegraphics[width=0.75\textwidth]{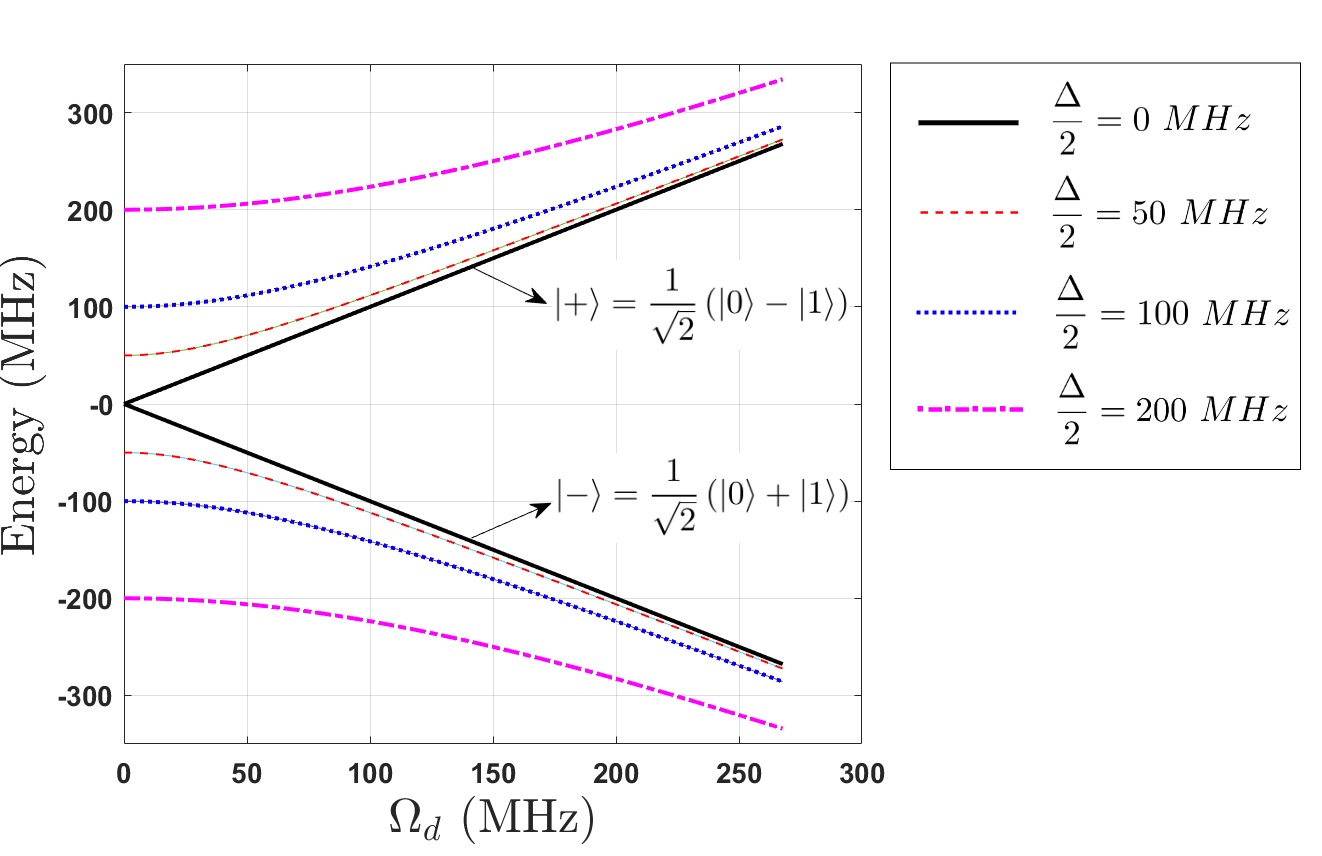}
	\caption{The variation of energy eigenvalues of the Hamiltonian, $H_{eff}$ when the value of $\Omega_{d}$ adiabatically increased from $0$ to $267.77 MHz$. The variation of energy eigenvalues are plotted for different detuning values, $\Delta$.}
\label{EnergyLevelDiagram}
\end{figure}

\begin{figure}[H]
\centering
\begin{subfigure}{0.49\textwidth}
\includegraphics[width=\textwidth]{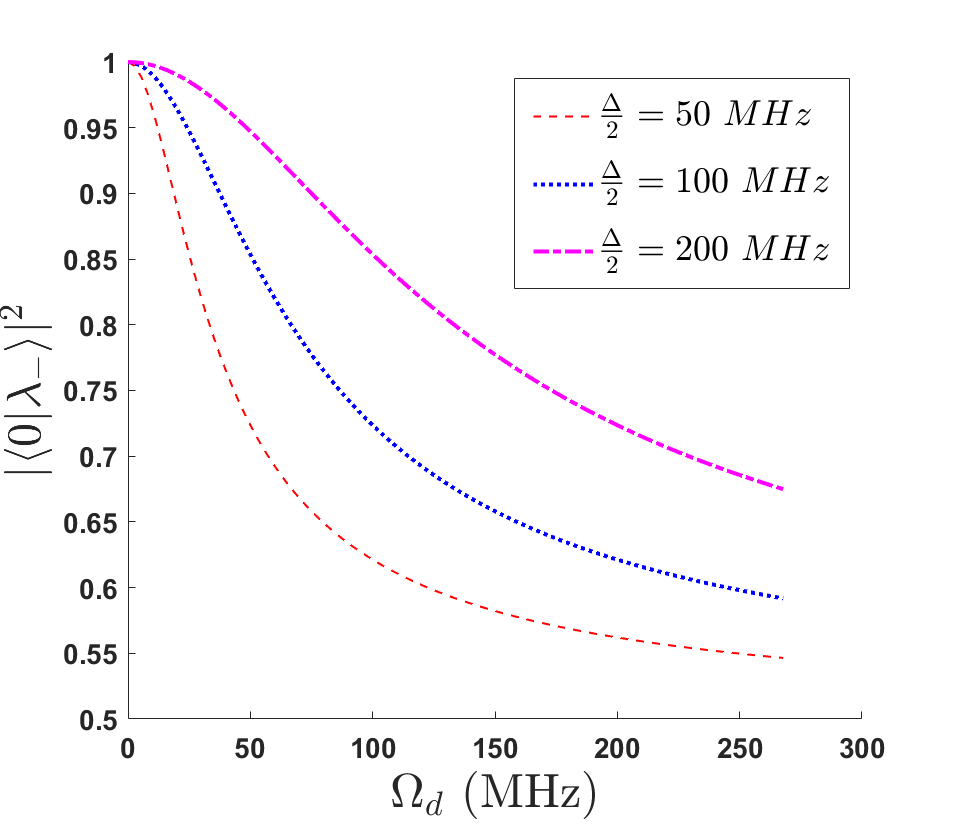}
\caption{}
\label{AdiabaticEvoOfGroundState}
\end{subfigure}
\begin{subfigure}{0.49\textwidth}
\includegraphics[width=\textwidth]{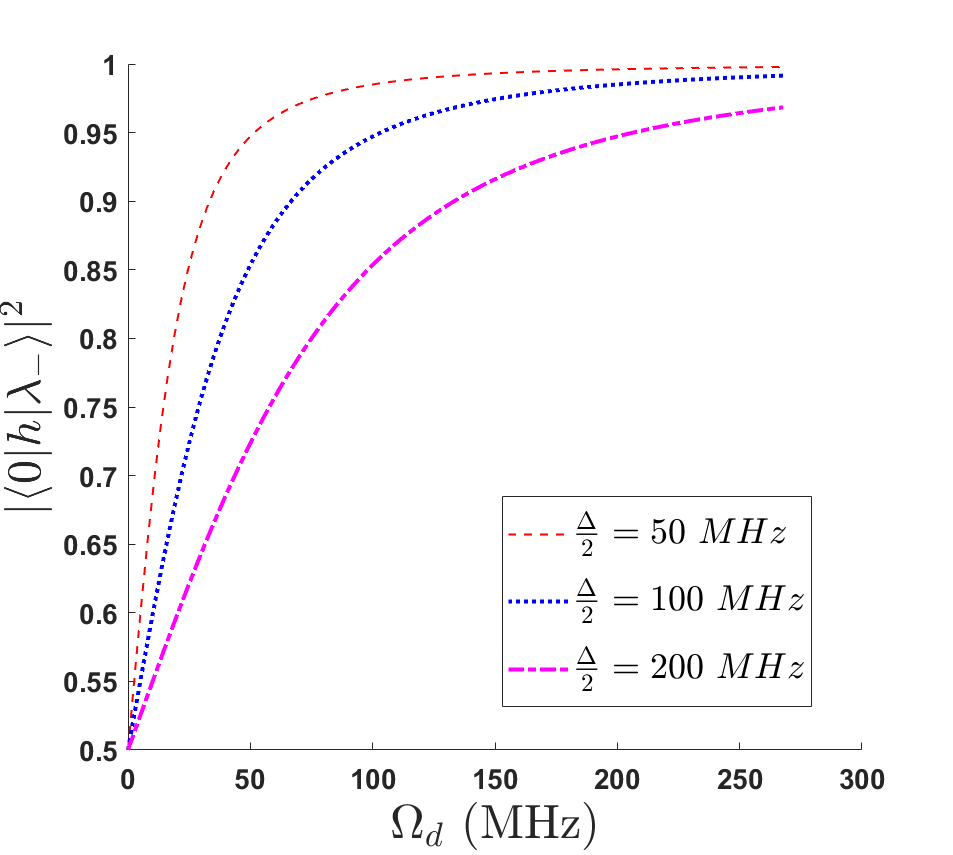}
\caption{}
\label{HadamardDiscriAdia}
\end{subfigure}
\end{figure}

We begin this section by explaining the adiabatic dynamics of a superconducting qubit controlled by electromagnetic pulses~\cite{KrantzA2019}. \autoref{Qubitpulse} is a schematic diagram of a qubit controlled by electromagnetic pulses.
The two energy levels of the qubit are separated by an energy gap corresponding to the qubit frequency, $\omega_{c}$ and the control pulse are propagated at a driving frequency of $\omega_{d}$. An effective Hamiltonian to describe the qubit coupled to an electromagnetic drive line is
\begin{equation}
H_{eff}=-\frac{1}{2}\Delta \sigma_{z}-\Omega_{d}\sigma_{x},
\label{QubitEffHamil}
\end{equation}
where $\Delta=\omega_{q}-\omega_{d}$ is the detuning and $\Omega_{d}=d(t)\Omega_{c}$ is the time-dependent amplitude of the control pulses. We assume the time-dependent parameter $d(t)$ can take values from the range $\left[0,1\right]$ resulting in pulses of time-dependent amplitude ranging from $0$ to $\Omega_{c}$, where $\Omega_{c}$ is a constant~\cite{DavidQiskit2018,ThomasQiskit2020}. Comparing with the general Hamiltonian of the a two level system given in \autoref{HamOfGenQub}, we can write,
\begin{equation}
\begin{split}
x&=-\Omega_{d}\\
y&=0\\
z&=-\frac{1}{2}\Delta. 
\end{split}
\end{equation}
Using a fixed frequency qubit ($\omega_{c}$ is a constant) and a fixed frequency driving pulse ($\omega_{d}$ is a constant), the function $z$ becomes time-independent and we are left with only one time-dependent parameter $x=-\Omega_{d}$. The eigenvalues of the Hamiltonian in this case are $\lambda_{\pm}(t)=\pm f(t)=\pm\sqrt{\left(\frac{\Delta}{2}\right)^{2}+\Omega_{d}^{2}}$ and the energy eigenstates are $\vert \lambda_{-}\rangle$ and $\vert \lambda_{+}\rangle$. \autoref{EnergyLevelDiagram} shows the variation of the energy eigenvalues as the value of $\Omega_{d}$ adiabatically (slowly) increases from $0$ to $267.77 MHz$. 
 The solid black line represents the resonant driving of the qubit ($\omega_{q}=\omega_{d}$), where the Hamiltonian will have only the Pauli-X component.Thus the state of the qubit is expected to be the eigenstates of $\sigma_{x}$, $\vert -\rangle=\frac{1}{\sqrt{2}}\left(\vert 0\rangle+\vert 1\rangle\right)$ and $\vert +\rangle=\frac{1}{\sqrt{2}}\left(\vert 0\rangle-\vert 1\rangle\right)$, where $\vert 0\rangle$ and $\vert 1\rangle$ are the computational basis (the eigenstates of Pauli-Z operator). We have also plotted the change in the energy eigenvalues with detuned driving with $\frac{\Delta}{2}=50~MHz$ (dashed red curve), $\frac{\Delta}{2}=100~MHz$ (dotted blue curve), and $\frac{\Delta}{2}=200~MHz$ (dash-dotted pink curve).  For all the curves with detuned driving, the energy eigenvalues show avoided crossing at $\Omega_{d}=0$ and approaches towards black line as $\Omega_{d}$ increases. Thus, If we are starting a process from $\Omega_{d}=0$ (the Hamiltonian will consist of only the $\sigma_{z}$ component) and from any one of the eigenstates of the Hamiltonian ($\vert 0\rangle$ or $\vert 1\rangle$), we can expect an adiabatic final state, which is closer to eigenstates of $\sigma_{x}$ for large adiabatic change in the amplitude of the pulse. For the detuned driving of the qubit, the adiabatic evolution of the ground state of the system Hamiltonian, $\vert \lambda_{-}\rangle$ with change in the amplitude of the drive $\Omega_{d}$ is shown in \autoref{AdiabaticEvoOfGroundState}. We have plotted the probability of occupation of $\vert \lambda_{-}\rangle$ to be in the computational ground state, $\vert \langle 0\vert \lambda_{-}\rangle\vert^{2}$ to show the adiabatic evolution (Throughout this paper, we will use the probability of the evolved states to be in the computational ground state as we can experimentally verify it using the modern day quantum computers). The probability of $\vert \lambda_{-}\rangle$ to be in the $\vert 0\rangle$ is unity when $\Omega_{d}=0$. Thus, we assume the initial state of the qubit is always the computational ground state irrespective of the detuning of the driving pulse. The approach of the adiabatic evolution of $\vert \lambda_{-}\rangle$ towards the $\vert -\rangle$ is evident from \autoref{AdiabaticEvoOfGroundState} as the probability curves declines and approach to $\vert\langle 0\vert \lambda_{-}\rangle\vert^{2}=0.5$ with increasing value of pulse amplitude.
Although, the $\vert \langle 0\vert \lambda_{-}\rangle\vert^{2}$ curve is not sufficient to verify the adiabaticity of the evolved state as both of the states $\vert -\rangle$ and $\vert +\rangle$ have $0.5$ probability to be in the computational ground state. To verify the adiabaticity, we need to confirm the closeness of the evolved state to the $\vert -\rangle$ state and it is possible to use Hadamard discrimination for the same. A Hadamard discrimination is obtained by applying a Hadamard gate ($h$) to the evolved state and finding the probability of the resultant state ($h\vert \lambda_{-}\rangle$)  to be in the computational zero state ($\vert\langle 0\vert h\vert \lambda_{-}\rangle\vert^{2}$). As we always begin the process from $\vert 0\rangle$ state, the value of the Hadamard discrimination should be close to $1$ for an adiabatically evolved state and considerably large changes in $\Omega_{d}$. The result of Hadamard discrimination for the change of $\Omega_{d}$ from $0$ to $267.77~MHz$ is shown in \autoref{HadamardDiscriAdia}. As we expected, the adiabatically evolved state approaches to $\vert -\rangle$ state for large change in $\Omega_{d}$ and it is indicated by the approach of the Hadamard discrimination values to unity.  

\subsection*{Non-adiabatic dynamics}
\label{NonAdiabaticDynamics}
\begin{figure}[H]
\centering
\begin{subfigure}{0.49\textwidth}
\includegraphics[width=\textwidth]{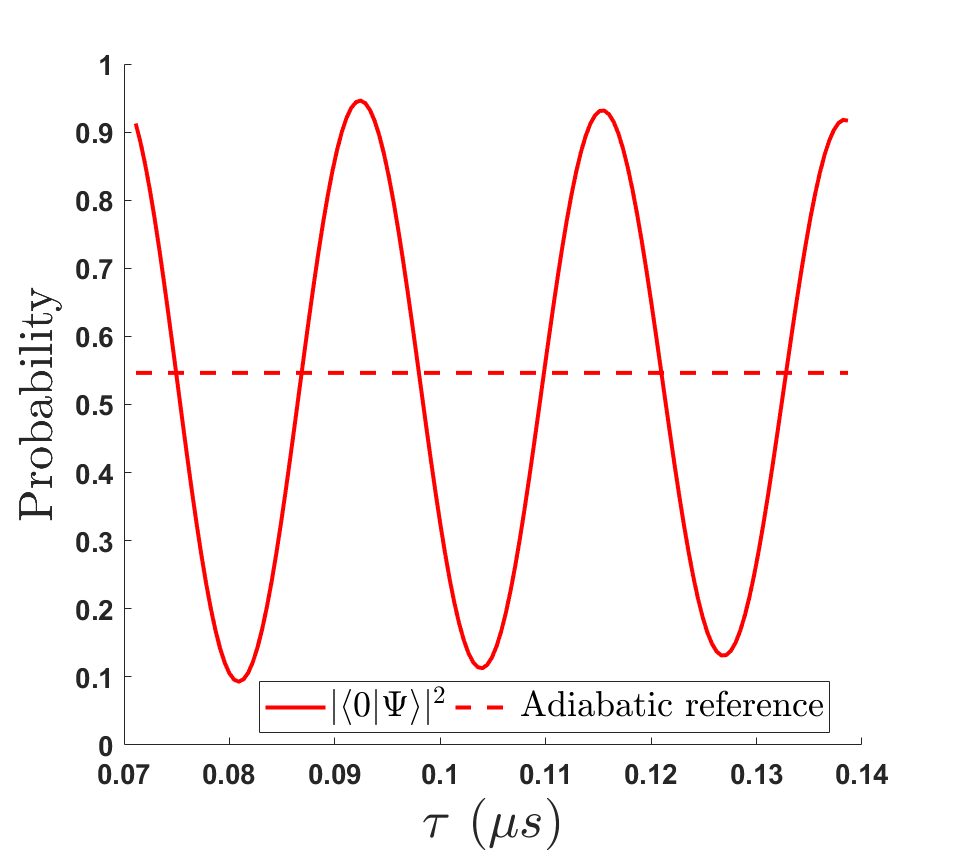}
\caption{}
\label{NonAdiabaticTheo05}
\end{subfigure}
\begin{subfigure}{0.49\textwidth}
\includegraphics[width=\textwidth]{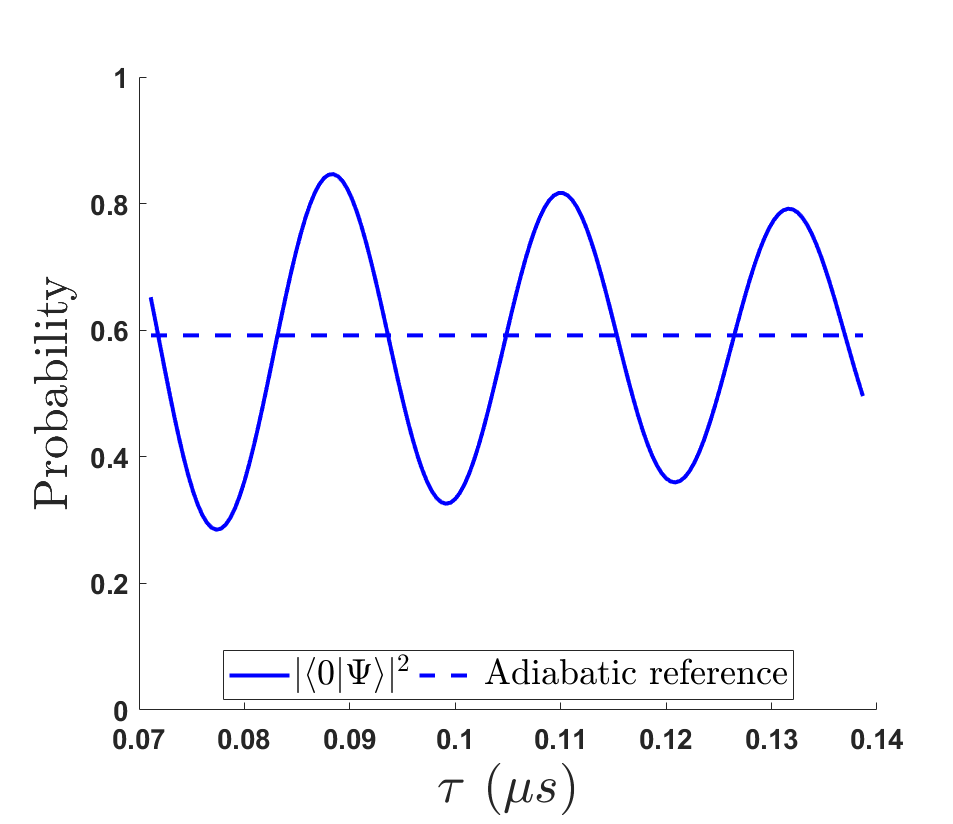}
\caption{}
\label{NonAdiabaticTheo1}
\end{subfigure}
\begin{subfigure}{0.49\textwidth}
\includegraphics[width=\textwidth]{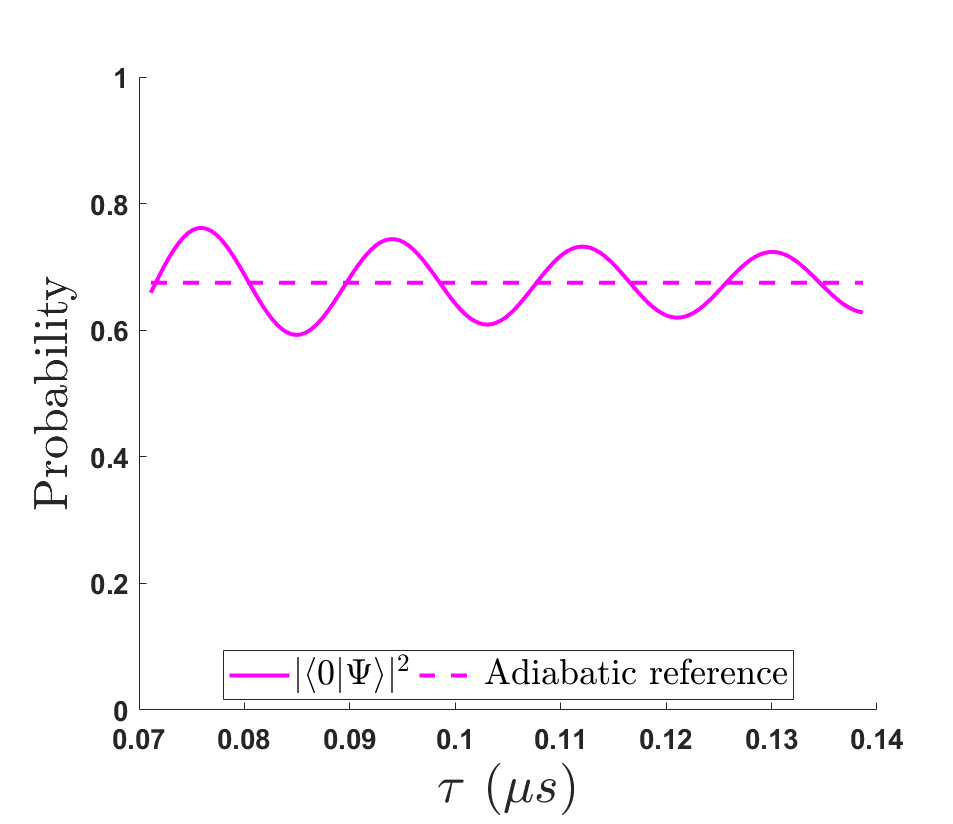}
\caption{}
\label{NonAdiabaticTheo2}
\end{subfigure}
\caption{}
\label{NonAdiabaticTheo}
\end{figure}

The finite time evolution of the qubit governed by $H_{eff}$ is described by the time-dependent Schrodinger equation,
\begin{equation}
i\frac{\partial}{\partial t}\vert \Psi \rangle=H_{eff}\vert \Psi \rangle,
\label{TDSEHeff}
\end{equation}
where $\vert \Psi\rangle$ is the time-dependent evolved state. Setting the initial state of the qubit as the computational ground state, $\vert 0\rangle$, the numerical calculation of the solution $\vert \Psi\rangle$ is possible using the equation,
\begin{equation}
\vert\Psi\rangle=\mathcal{T}exp\left(-i\int_{0}^{\tau}H_{eff}~dt\right)\vert 0\rangle,
\end{equation}
where $\mathcal{T}$ represents the time ordering and $\tau$ is the time duration of the evolution. The non-adiabatic evolution of the qubit governed by the $H_{eff}$ is numerically calculated for a linearly increasing value of $\Omega_{d}$, i.e., The value of $\Omega_{d}$ linearly increases from $0$ to $267.77~MHz$ in $\tau$ time,
\begin{equation}
\Omega_{d}=267.77\left(\frac{t}{\tau}\right),
\end{equation}
where time $t$ varies from $0$ to $\tau$. The probability of the obtained evolved state, $\vert \Psi\rangle$, to be in the computational ground state is plotted against the time duration $\tau$ of the process in \autoref{NonAdiabaticTheo}.
The value of $\vert\langle 0\vert\Psi\rangle\vert^{2}$ oscillates about the adiabatic reference probability and it will converge completely to the adiabatic probability asymptotic with time ($\tau=\infty$) according to the quantum adiabatic theorem. It is also evident from the plot that the time duration $\tau$ required for $\vert\langle 0\vert\Psi\rangle\vert^{2}$ to converge is less for large values of $\frac{\Delta}{2}$ than that of the small values of $\frac{\Delta}{2}$. Thus, a non-adiabatic drive of the qubit with highly detuned pulse can attain the adiabatic final state in considerably less time duration $\tau$. However, it is important to notice that the adiabatic reference probability approaches unity with highly detuned pulses, i.e., the qubit will remain in its initial state, $\vert 0\rangle$, throughout the evolution as the effect of the $\sigma_{x}$ component is negligible compared with large detuning values associated with the $\sigma_{z}$ component of the Hamiltonian. Thus, a considerable adiabatic change in the state of the qubit is attainable only with processes of large time duration.   

The application of NOBIE method in this case is expected to drive the qubit achieve the adiabatic final state irrespective the time duration of the process. In the following subsection, we derive the NOBIE Hamiltonian for the qubit controlled by electromagnetic pulse.
 
\subsection*{NOBIE Hamiltonian}
The number operator for the Hamiltonian in \autoref{QubitEffHamil} can be obtained from the general \autoref{NumOpeForTwoLev} as
\begin{equation}
\mathcal{N}(t)=\frac{-\Omega_{d}\sigma_{x}-\frac{1}{2}\Delta\sigma_{z}}{\sqrt{\left(\frac{\Delta}{2}\right)^{2}+\Omega_{d}^{2}}}
\label{NumOpeForHeff}
\end{equation}
The parameters of the NOBIE Hamiltonian for the above qubit coupled with electromagnetic pulse can be found using \autoref{CDHamiltonian} and \autoref{IndependentSolutions}. The NOBIE Hamiltonian corresponding to the mutually dependent solutions is
\begin{equation}
H_{N}=H_{eff}+\frac{\Delta\dot{\Omega}_{d}}{4f^{2}(t)}\sigma_{y},
\label{MutuHN}
\end{equation}
and the NOBIE Hamiltonian corresponding to the mutully independent solutions is
\begin{equation}
H_{N}=\frac{\Delta\dot{\Omega}_{d}}{4f^{2}(t)}\sigma_{y}.
\label{MutuInHN}
\end{equation}
The $\sigma_{y}$ component of the above NOBIE Hamiltonians are equivalent to each other, which makes the presence of the additional system Hamiltonian in \autoref{MutuHN} irrelevant to attain shortcuts to adiabaticity. The above results are supported by the fact that addition or subatraction of the system Hamiltonian with the NOBIE Hamiltonian will result in a Hamiltonian that can satisfy the invariance condition in \autoref{NumberInvariance}. Although, both the above NOBIE Hamiltonians can drive the qubit to obtain shortcut to adiabaticity, the $H_{N}$ corresponding to the mutually independent solutions need only one control parameter for the implementation while the other needs all the three controls. The lower number of control parameters motivates us to use the NOBIE Hamiltonian in \autoref{MutuInHN} for the realization of NOBIE method in real qubits. The experimental observations of non-adiabatic and shortcut driving of a real transmon qubit in an IBM quantum computer is provided in the following section.

\section{Experimental realization using IBM Quantum Lab} 
IBM quantum compurers use transmon qubits coupled to electromagnetic pulses for quantum computation. We can access the qubits of the quantum computers to execute quantum experiments using the online platform, IBM Quantum Lab. The experiments are designed using an open-source software development kit (SDK), Qiskit based on Python programming language. We have used the quantum computer, $ibm\_oslo$ to execute the single qubit non-adiabatic and NOBIE driving. $ibm\_oslo$ is a $7$ qubit backend, in which we chose the qubit $0$ for experiments. We work with the calibrated frequency of the qubit, $\omega_{q}=4.925035720219493\times 2\pi~GHz$ and the frequency of the electromagnetic pulse ($\omega_{d}$) is assigned according to the required duruning, $\Delta$. We use Qiskit Pulse package to give pulse level instructions to the backends and assign the time-dependent pulse amplitude by setting the values of $d(t)\in[0,1]$, where $d(t)=1$ gives the maximum attainable amplitude, $\Omega_{c}$. The drive line of the selected qubit has $\Omega_{c}=535.54~MHz$, which is calculated using the Rabi oscillation of the qubit state by applying the resonant constant pulses (see \nameref{Appendix B} for complete calculation). The calculated value of $\Omega_{c}$ might have some deviation from the actual value of maximum attainable pulse amplitude. This deviation is due to the use of an effective Hamiltonian (\autoref{QubitEffHamil}) which assumes the qubit is not connected to the other qubits. The calculated value of $\Omega_{c}$ might result in small deviation of experimental outputs from the expected theoretical predictions. However, $\Omega_{c}=535.54~MHz$ is enough for a qualitative analysis of both non-adiabatic and NOBIE methods. The minimum time duration to assign the pulse amplitude (sampling time) of the drive line is $dt=0.22222222222222221~ns$. Thus, we can only assign step by step variation of pulse amplitude instead of the continuous variation required by the theoretical predictions. The qubit resets to the computational ground state, $\vert 0\rangle$ before the execution of designed experiments. In all the experiments, we measure the probability of the evolved state to be in the $\vert 0\rangle$ state. We use $1024$ shots for a specified drive, and repeat a single set of experiment $30$ times to obtain the results.
\subsection*{Non-adiabatic driving}
\begin{figure}[H]
\centering
\begin{subfigure}{0.49\textwidth}
\includegraphics[width=\textwidth]{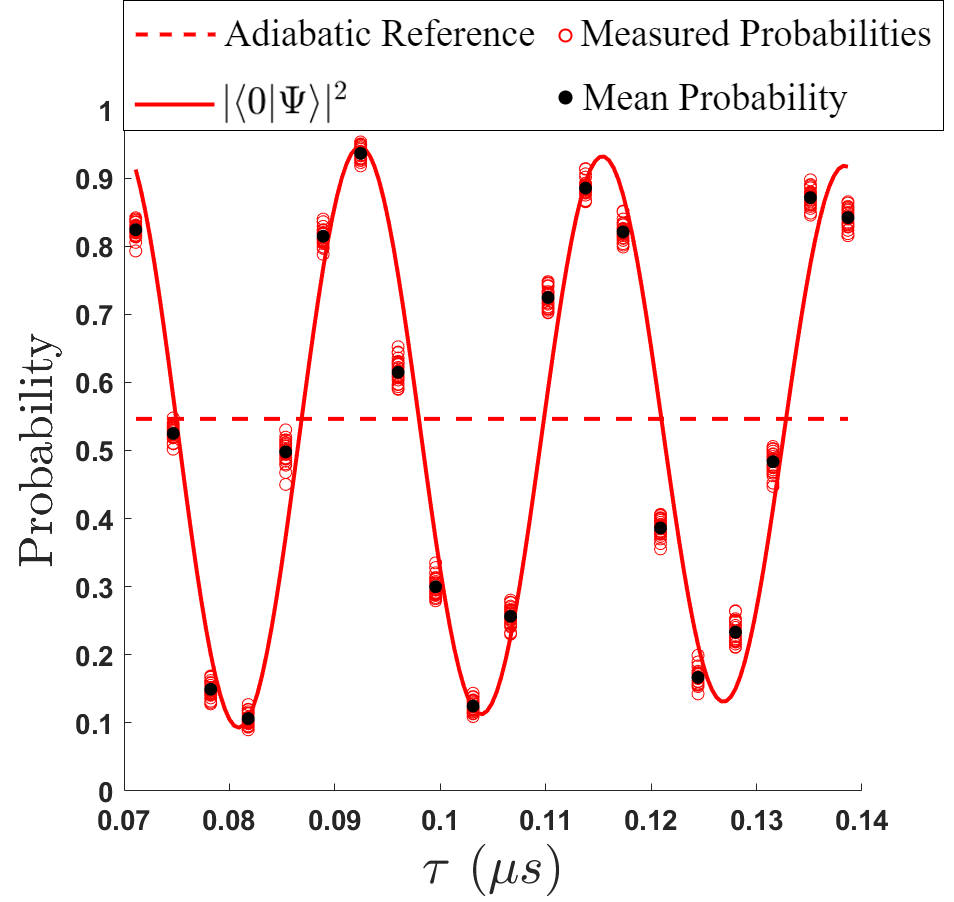}
\caption{}
\label{NonAdiaExpTheoComb05}
\end{subfigure}
\begin{subfigure}{0.49\textwidth}
\includegraphics[width=\textwidth]{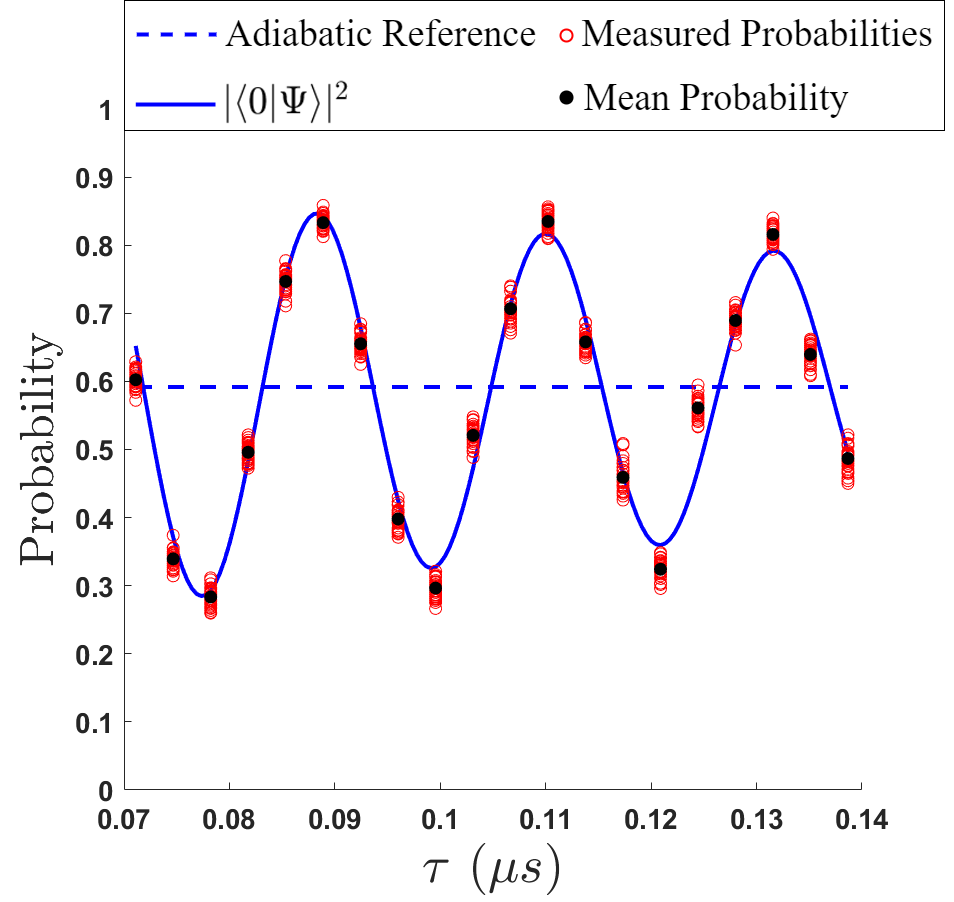}
\caption{}
\label{NonAdiaExpTheoComb1}
\end{subfigure}
\begin{subfigure}{0.49\textwidth}
\includegraphics[width=\textwidth]{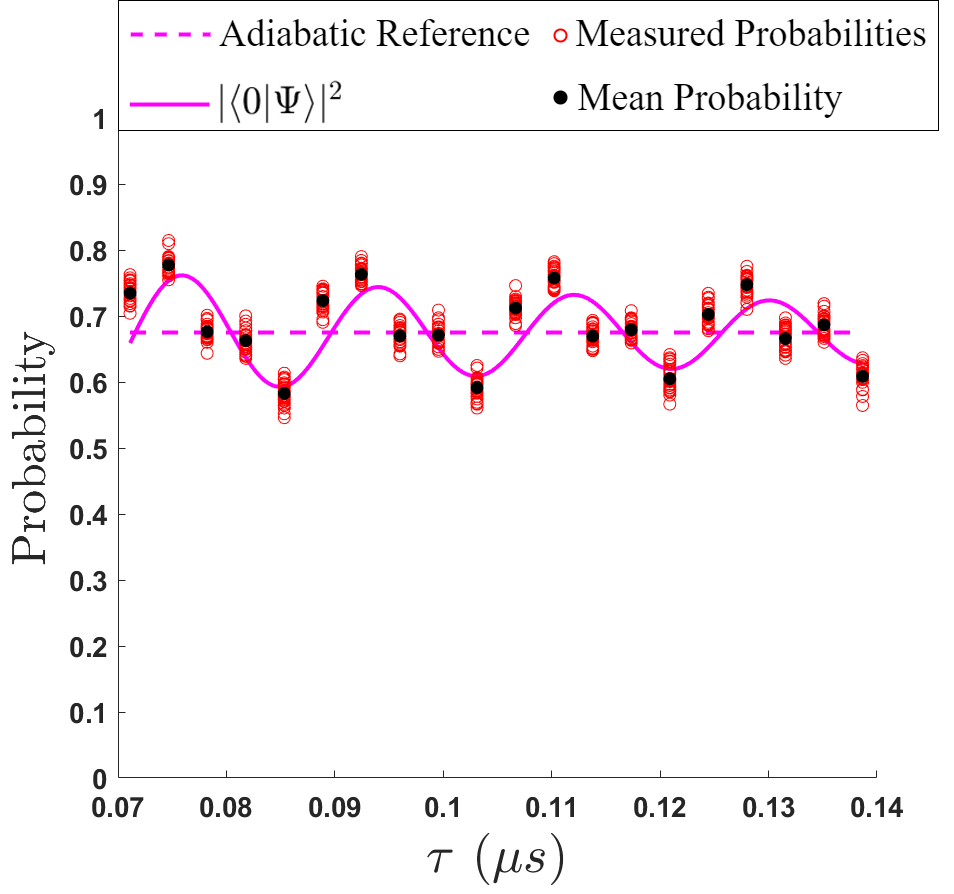}
\caption{}
\label{NonAdiaExpTheoComb2}
\end{subfigure}
\caption{}
\label{NonAdiaExpTheoComb}
\end{figure}

The non-adiabatic driving explained in \autoref{NonAdiabaticDynamics} requires to linearly increase the value of $\Omega_{d}=d(t)\Omega_{c}$ from $0$ to $267.77~MHz$. Considering the calculated value of $\Omega_{c}$, we need to linearly increase the value of $d(t)$ from $0$ to $0.5$ as,
\begin{equation}
d(t)=0.5\left(\frac{t}{\tau}\right).
\end{equation}
We have executed the finite driving of the qubit under the governance of the Hamiltonian in \autoref{QubitEffHamil} for various time durations $\tau$. We have applied in-phase pulses to attach the linear change in the pulse amplitude to the $\sigma_{x}$ dynamics. The complete Python program is given below (see \autoref{code1}).
\begin{lstlisting}[caption= Python code for non-adiabatic drive of the qubit, label=code1, style=chstyle, language=Python]
#Access the IBM Quantum Lab account, provider, and the backend

import warnings
warnings.filterwarnings('ignore')
from qiskit.tools.jupyter import *
from qiskit import IBMQ
IBMQ.load_account()
provider = IBMQ.get_provider(hub='ibm-q-research-2', group='central-uni-tami-1', project='main')
backend = provider.get_backend('ibm_oslo')

#Import all the necessary packages

from qiskit.tools.monitor import job_monitor
from qiskit import schedule
import numpy as np
from qiskit import pulse     # importing pulse package    
from qiskit.circuit import QuantumCircuit, Gate

backend_config = backend.configuration()
dt = backend_config.dt #defining the sampling time
backend_defaults = backend.defaults()
qubit = 0  #to select qubit zero for experiments
center_frequency_Hz = backend_defaults.qubit_freq_est[qubit]        #Loading the frequency of the qubit in in Hz


drive_durations=[320, 336, 352, 368, 384, 400, 416, 432, 448, 464, 480, 496, 512, 528, 544, 560, 576, 592, 608, 624] #drive durations defined as number of samples and it should be multiples of 16
for drive_duration in drive_durations:

    qc_circs=[] #Initializing list for appending quantum circuits
    with pulse.build(backend=backend, default_alignment='sequential', name='Non-adiabatic Experiment') as Non_sched:
            drive_chan = pulse.drive_channel(qubit)   #selecting drive channel
            pulse.set_frequency(center_frequency_Hz-((1*(10**8))/(2*np.pi)), drive_chan) #setting the frequency of the pulse
            t0 = 0   # setting initial time, t=0
            tf = drive_duration*dt  #setting the time duration of the process in seconds
            intervals =  np.arange(t0, tf, dt).tolist()  #spliting the total time durations into intervals of sampling time
            drive_amps = [((2*j)+dt)/(2*2*tf) for j in intervals]  #defining linear increase of the d(t) from 0 to 0.5
            pulse.play(drive_amps, drive_chan)  #command to play the pulse

    Non_gate = Gate("Non", 1,[])     #setting a custom gate to execute non-adiabatic drive
    qc_Non = QuantumCircuit(1, 1)  #defining the quantum circuit the circuit
    qc_Non.append(Non_gate, [0])   #appending gates to circuit
    qc_Non.measure(0, 0)  #command to measure the qubit
    qc_Non.add_calibration(Non_gate, (0,), Non_sched, [])  #adding the Non_sched as a calibration to the custom gate
    S=30  #defining the number to repeat the experiments 
    for i in range(S):
        qc_circs.append(qc_Non) #appending the quantum circuit for repeating experiments
    num_shots_per_dur = 1024 #defining total number of shots

    job = backend.run(qc_circs, 
                  meas_level=2, 
                  meas_return='single', 
                  shots=num_shots_per_dur)  #assigning job to the backend

    job_monitor(job) #monitor the status of the assigned job
    Non_results = job.result(timeout=120) #acquiring the results of the experiments
    prob0=[] #initializing the list to store the probabilities measured
    for k in range(S):
        prob0.append((Non_results.get_counts(k)['0'])/1024) #appending the results of 30 experiments
    #Print the results
    
    print(drive_duration) #time durations of the non-adiabatic process as number of samples
    print(prob0)  #measured probabilities
    print(np.mean(prob0))  #mean value of measured probabilities                                                                  
\end{lstlisting}
The given \autoref{code1} is written for the detuning value $\Delta=100~MHz$ (see line 32 of the Python code). The time duration for each experiment is specified in terms of number of samples and it should be a multiple of $16$ to be compatible with the selected IBM quantum computer. We have increased the time duration of the process from $0.0711~\mu s$ ($320$ samples) to $0.13866~\mu s$ ($624$ samples) with step size of $0.00355~\mu s$ ($16$ samples). The variation of pulse amplitude for the time duration: $\tau=0.0711~\mu s$ and detuning: $\Delta=100 MHz$ is given in \autoref{linearPulse}. The linear increment of the pulse amplitude is defined in the line $36$ of \autoref{code1} and applied using $pulse.play$ instruction in line $37$. The $meas\_level$=2 in \autoref{code1} imples the state measurement of the qubit on completion of the drive to know whether the qubit is in $\vert 0\rangle$ state or $\vert 1\rangle$ state. A single experiment will execute the drive and measurements 1024 times (see line $47$ and $52$ in \autoref{code1}) and find the probability to be in $\vert 0\rangle$ state by dividing the total number of times we obtained $\vert 0\rangle$ by 1024 (see line $58$ of \autoref{code1}). The measured probabilities of such $30$ experiments and their mean value for various time durations are plotted in \autoref{NonAdiaExpTheoComb}. In the above figure, we have merged the experimental values with the curves of theoretical predition for comparison. The experimental observation closely follows the numerical results proving the non-adiabaticity in the drive using pulses of linearly increasing amplitude. The measured probabilities are vertically spread for each time duration showing the error in evolution of the qubit. The small deviation of the mean probability from the predicted values is expected because of two reasons. The first reason is the consideration of the effective Hamiltonian assuming the qubit is isolated from other qubit, but it is not in real quantum computer. Thus, interactions from the other qubits might cause deviation from the expected output. Secondly, the error in the measurement is acceptable as we are using a NISQ (Noisy Intermediate-Scale Quantum) device. The pulse we used might be noisy enough to deviate from the expected evolution of the qubit.

In the following subsection, we execute the evolution of the qubit under the NOBIE Hamiltonian to attain the adiabatic state irrespective of the time duration of the process. The Hadamard discrimination is also performed to verify the adiabaticity of the evolved state under shortcut evolution.

\subsection*{NOBIE driving}
\begin{figure}[H]
\centering
\begin{subfigure}{0.49\textwidth}
\includegraphics[width=\textwidth]{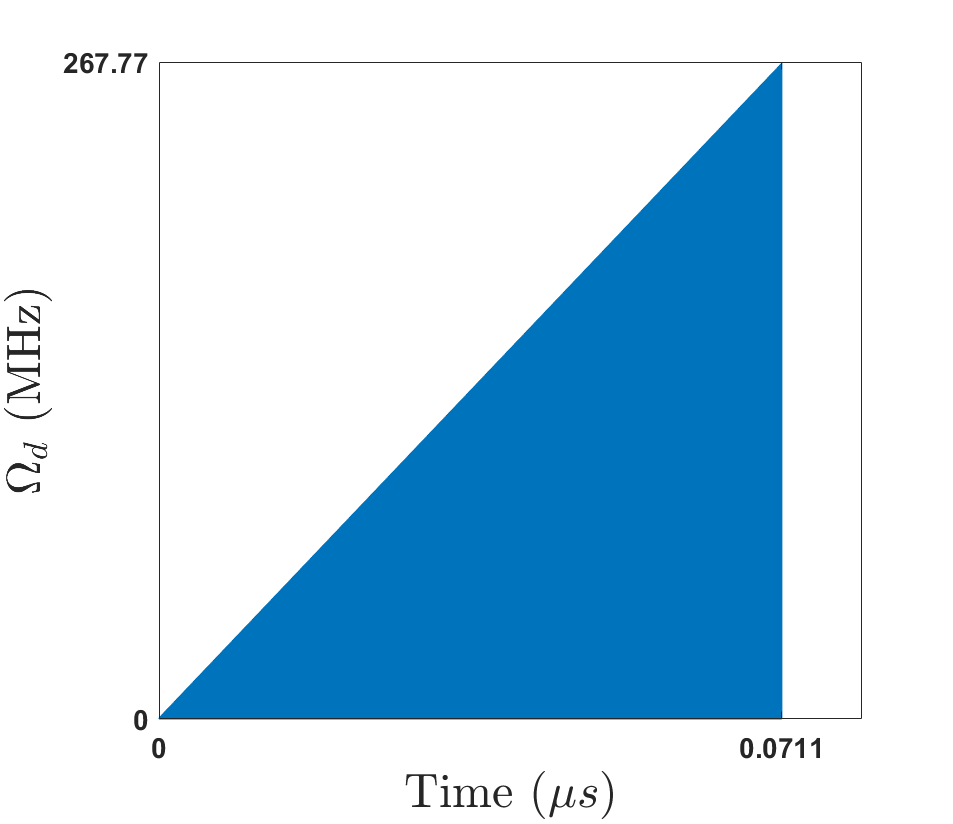}
\caption{}
\label{linearPulse}
\end{subfigure}
\begin{subfigure}{0.49\textwidth}
\includegraphics[width=\textwidth]{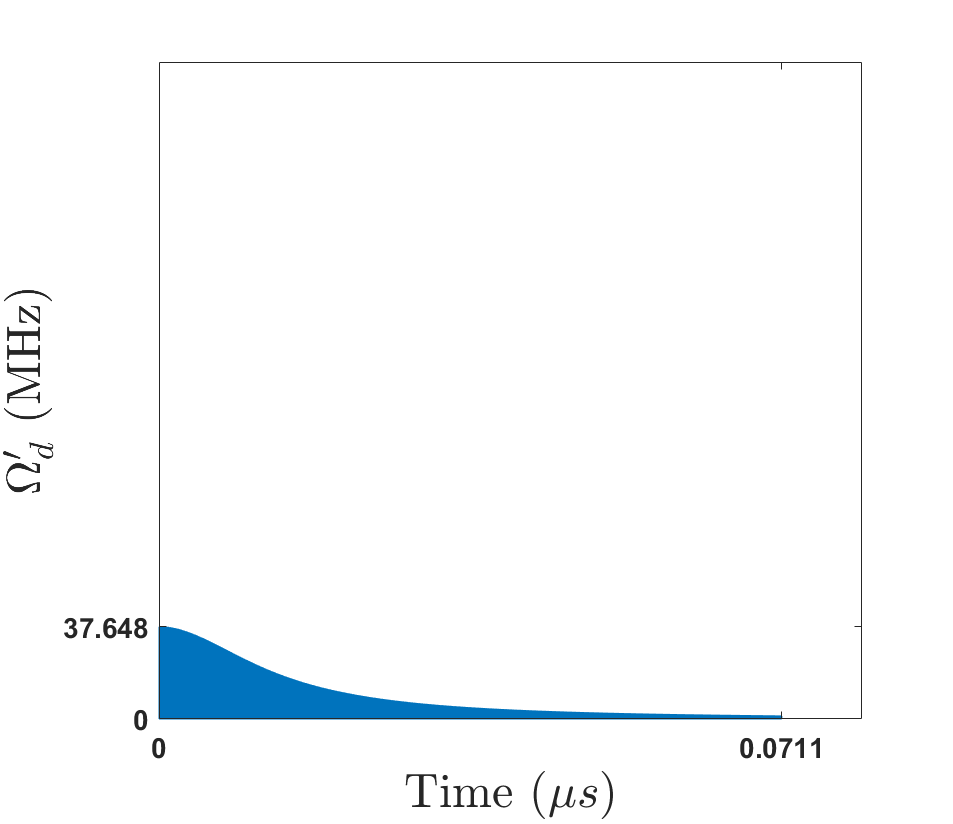}
\caption{}
\label{NOBIEpulse}
\end{subfigure}
\caption{}
\label{PulsesDiagram}
\end{figure}

The NOBIE Hamiltonian in \autoref{MutuInHN} can be mapped to an effective Hamiltonian of the form,
\begin{equation}
H_{N}=\frac{1}{2}\Delta^{\prime} \sigma_{z}+\Omega^{\prime}_{d}\sigma_{y},
\label{NOBIEqubHamil}
\end{equation}
where the detuning of driving pulse, $\Delta^{\prime}=0$ (we need to apply resonant pulses) and the pulse amplitude, $\Omega^{\prime}_{d}=\frac{\Delta\dot{\Omega}_{d}}{4f^{2}(t)}$. The above variation of pulse amplitude for the time duration: $\tau=0.0711~\mu s$ is given in \autoref{NOBIEpulse}, which can be assigned to quantum computer as a list of amplitude values for each sample of time duration $dt$. The phase of the pulse should be fixed as $\frac{\pi}{2}$ associate the pulse variation to the $\sigma_{y}$ component of the Hamiltonian. The complete python program to run the NOBIE experiments is given below.

\begin{figure}[H]
\centering
\begin{subfigure}{0.49\textwidth}
\includegraphics[width=\textwidth]{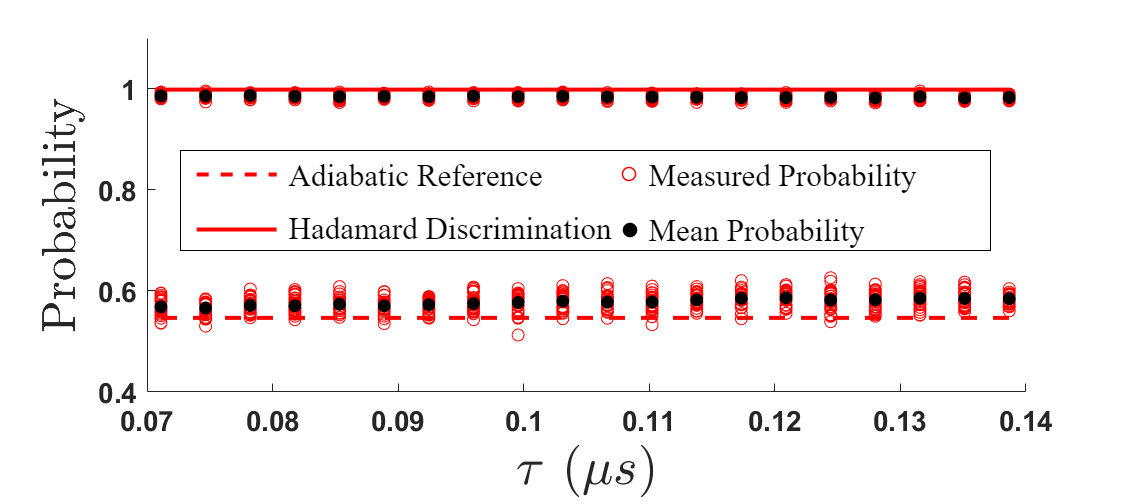}
\caption{}
\label{NOBIE_Exp_Theo_Comb_05108}
\end{subfigure}
\begin{subfigure}{0.49\textwidth}
\includegraphics[width=\textwidth]{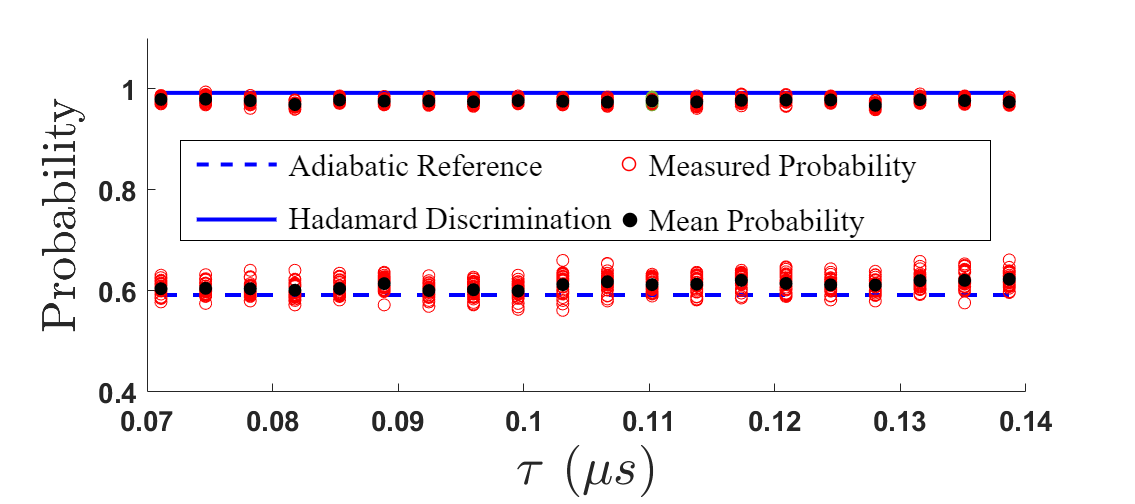}
\caption{}
\label{NOBIE_Exp_Theo_Comb_1108}
\end{subfigure}
\begin{subfigure}{0.49\textwidth}
\includegraphics[width=\textwidth]{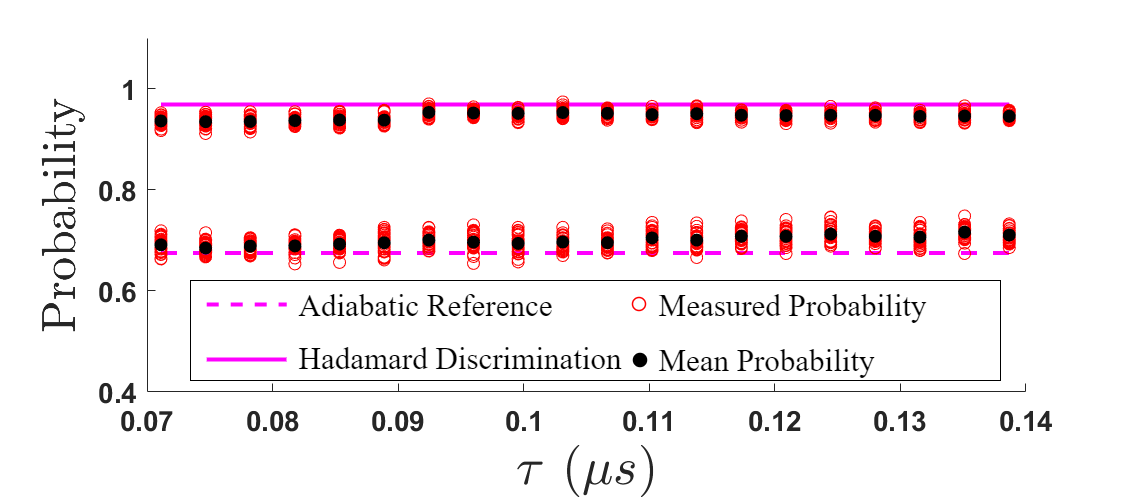}
\caption{}
\label{NOBIE_Exp_Theo_Comb_2108}
\end{subfigure}
\caption{}
\label{NOBIE_Exp_Theo_Comb}
\end{figure}

\begin{lstlisting}[caption= Python code for NOBIE drive of the qubit, label=code2, style=chstyle, language=Python]
#Access the IBM Quantum Lab account, provider, and the backend

import warnings
warnings.filterwarnings('ignore')
from qiskit.tools.jupyter import *
from qiskit import IBMQ
IBMQ.load_account()
provider = IBMQ.get_provider(hub='ibm-q-research-2', group='central-uni-tami-1', project='main')
backend = provider.get_backend('ibm_oslo')

#Import all the necessary packages
from qiskit import transpile
from qiskit.tools.monitor import job_monitor
from qiskit import schedule
import numpy as np
from qiskit import pulse     # importing pulse package    
from qiskit.circuit import QuantumCircuit, Gate

backend_config = backend.configuration()
dt = backend_config.dt #defining the sampling time
backend_defaults = backend.defaults()
qubit = 0  #to select qubit zero for experiments
center_frequency_Hz = backend_defaults.qubit_freq_est[qubit]        #Loading the frequency of the qubit in in Hz


drive_duration=320 #drive duration defined as number of samples and it should be a multiple of 16
qc_circs=[] #Initializing list for appending quantum circuits
with pulse.build(backend=backend, default_alignment='sequential', name='Non-adiabatic Experiment') as Non_sched:
    drive_chan = pulse.drive_channel(qubit)   #selecting drive channel
    pulse.set_frequency(center_frequency_Hz, drive_chan) #setting the frequency of the pulse
    drive_amps=[0.070308,0.070268,0.07019,0.070072,0.069916,0.069722,
      0.06949,0.069222,0.068918,0.068579,0.068206,0.067801,0.067364,
      0.066898,0.066402,0.06588,0.065331,0.064758,0.064162,0.063545,
      0.062908,0.062253,0.061581,0.060894,0.060193,0.05948,0.058756,
      0.058023,0.057281,0.056533,0.055779,0.055021,0.05426,0.053497,
      0.052733,0.051969,0.051206,0.050444,0.049686,0.04893,0.048179,
      0.047432,0.046691,0.045956,0.045228,0.044506,0.043792,0.043085,
      0.042387,0.041697,0.041016,0.040344,0.03968,0.039026,0.038382,
      0.037747,0.037122,0.036507,0.035901,0.035305,0.034719,0.034143,
      0.033577,0.033021,0.032474,0.031937,0.031409,0.030891,0.030383,
      0.029884,0.029394,0.028913,0.028442,0.027979,0.027525,0.027079,
      0.026643,0.026214,0.025794,0.025382,0.024978,0.024582,0.024193,
      0.023812,0.023439,0.023072,0.022713,0.022361,0.022016,0.021678,
      0.021346,0.021021,0.020702,0.020389,0.020082,0.019782,0.019487,
      0.019198,0.018914,0.018636,0.018364,0.018096,0.017834,0.017577,
      0.017325,0.017077,0.016834,0.016596,0.016362,0.016133,0.015908,
      0.015688,0.015471,0.015259,0.01505,0.014845,0.014644,0.014447,
      0.014254,0.014064,0.013877,0.013694,0.013514,0.013337,0.013164,
      0.012994,0.012826,0.012662,0.012501,0.012342,0.012186,0.012033,
      0.011883,0.011735,0.01159,0.011447,0.011307,0.011169,0.011034,
      0.0109,0.010769,0.010641,0.010514,0.01039,0.010267,0.010147,
      0.010029,0.0099122,0.0097977,0.0096851,0.0095743,0.0094652,
      0.009358,0.0092524,0.0091486,0.0090464,0.0089458,0.0088468,
      0.0087494,0.0086535,0.0085591,0.0084661,0.0083746,0.0082845,
      0.0081958,0.0081085,0.0080224,0.0079377,0.0078543,0.0077721,
      0.0076911,0.0076114,0.0075328,0.0074554,0.0073791,0.007304,
      0.00723,0.007157,0.0070851,0.0070142,0.0069444,0.0068755,
      0.0068077,0.0067407,0.0066748,0.0066098,0.0065457,0.0064824,
      0.0064201,0.0063586,0.006298,0.0062382,0.0061793,0.0061211,
      0.0060638,0.0060072,0.0059513,0.0058963,0.0058419,0.0057883,
      0.0057354,0.0056832,0.0056317,0.0055809,0.0055307,0.0054812,
      0.0054324,0.0053841,0.0053365,0.0052895,0.0052431,0.0051973,
      0.0051521,0.0051074,0.0050633,0.0050198,0.0049768,0.0049344,
      0.0048924,0.004851,0.0048101,0.0047697,0.0047298,0.0046904,
      0.0046515,0.004613,0.004575,0.0045375,0.0045004,0.0044637,
      0.0044275,0.0043917,0.0043564,0.0043214,0.0042869,0.0042528,
      0.004219,0.0041857,0.0041527,0.0041201,0.0040879,0.0040561,
      0.0040246,0.0039935,0.0039628,0.0039323,0.0039023,0.0038725,
      0.0038431,0.0038141,0.0037853,0.0037569,0.0037287,0.0037009,
      0.0036734,0.0036462,0.0036193,0.0035927,0.0035663,0.0035403,
      0.0035145,0.003489,0.0034638,0.0034388,0.0034141,0.0033897,
      0.0033655,0.0033416,0.0033179,0.0032945,0.0032713,0.0032484,
      0.0032257,0.0032032,0.003181,0.0031589,0.0031372,0.0031156,
      0.0030942,0.0030731,0.0030522,0.0030315,0.003011,0.0029907,
      0.0029706,0.0029507,0.0029309,0.0029114,0.0028921,0.002873,
      0.002854,0.0028353,0.0028167,0.0027983,0.0027801,0.002762,
      0.0027442,0.0027265,0.0027089,0.0026916,0.0026744,0.0026573,
      0.0026404,0.0026237,0.0026071,0.0025907,0.0025745,0.0025584,
      0.0025424,0.0025266,0.0025109,0.0024954,0.00248,0.0024648,
      0.0024497,0.0024347,0.0024199,0.0024052,0.0023906,0.0023762]  #the list of amplitudes is manually substituted for each drive_duration  
    pulse.shift_phase(np.pi/2, drive_chan) #shifted the phase for y rotations
    pulse.play(drive_amps, drive_chan)  # command to play the pulse
    #pulse.shift_phase(-np.pi/2, drive_chan) #command to shift back the phase to switch of y rotations

Non_gate = Gate("Non", 1,[])     #setting a custom gate to execute non-adiabatic drive
qc_Non = QuantumCircuit(1, 1)  #defining the quantum circuit the circuit
qc_Non.add_calibration(Non_gate, (0,), Non_sched, [])  #adding the Non_sched as a calibration to the custom gate
qc_Non.append(Non_gate, [0])   #appending gates to circuit
#qc_Non.h(0) #command for hardamard operation
qc_Non.measure(0, 0)  #command to measure the qubit
qc_Non = transpile(qc_Non, backend) #transpiling to basis gates
S=30  #defining the number to repeat the experiments 
for i in range(S):
    qc_circs.append(qc_Non) #appending the quantum circuit for repeating experiments
num_shots_per_dur = 1024 #defining total number of shots

job = backend.run(qc_circs, 
                     meas_level=2, 
                     meas_return='single', 
                     shots=num_shots_per_dur)  #assigning job to the backend

job_monitor(job) #monitor the status of the assigned job
Non_results = job.result(timeout=120) #acquiring the results of the experiments
prob0=[] #initializing the list to store the probabilities measured
for k in range(S):
     prob0.append((Non_results.get_counts(k)['0'])/1024) #appending the results of 30 experiments
#Print the results
    
print(drive_duration) #time durations of the non-adiabatic process as number of samples
print(prob0)  #measured probabilities
print(np.mean(prob0))  #mean value of measured probabilities                                                                  
\end{lstlisting}
This python program works for a single $drive\_duration$ in single run and the corresponding list of pulse amplitudes (see lines $31$ to $80$ of \autoref{code2}) is manually substituted. The resonant drive is specified in line $30$ and the phase of the pulse is shifted $\frac{\pi}{2}$ in line $81$ to get the rotations around the y axis. We execute 30 experiments of 1024 shots for the NOBIE driving of a specified $drive\_duration$. Thus the result of the job (see line $97$ to $111$ of \autoref{code1}) gives the measured probabilty of the qubit to be in the $\vert 0\rangle$ state and their mean value. Uncomment the lines $83$ (shifts back the phase of the pulse to zero) and $89$ (Applying the Hadamard gate) to execute the Hadamard discrimination of the evolved qubit state. \autoref{NOBIE_Exp_Theo_Comb} shows the results of NOBIE driving and corresponding Hadamard discrimination. The theoretical adiabatic probability is shown using dashed lines for reference and the corresponding theoretical Hadamard discrimination probabilities are shown using solid lines. The measured probabilities are indicated using red circles and the mean value using black discs. Comparing with the probabilities of non-adiabatic driving, the probabilities of the NOBIE drive doesn't show any oscillation but gives values near to adiabatic probabilities irrespective of the time duration. The measured probability and the mean value show deviation from the theoretically predicted reference line as a result of experiment using a NISQ level quantum computer. The closeness of the measured and mean probabilities to the predicted Hadamard discrimination probability confirms the adiabatic state obtained using the NOBIE driving.

\section{Conclusion}
\label{Conclusion}
A method, NOBIE, for obtaining shortcuts to adibaticity through an entirely adiabatic path is developed for a general two level quantum system. The developed method gives two kinds of solutions, one with mutual dependence of the control parameters and the other with mutually independent control parameters. The mutually independent set of solutions provide an additional advantage of the reduced number of controls to obtain shortcut processes. Using the general formalism of NOBIE method, a shortcut protocol for a superconducting qubit driven by electromagnetic pulses is formulated. The experimental implementation of the formulated shortcut protocol is achieved in an IBM quantum computer using the IBM Quantum Lab. The evolution of a driven qubit in a quantum computer is measured as the probability to be in the computational ground state, $\vert 0\rangle$. Prior to the implementation of the shortcut protocol, the non-adiabatic dynamics of the qubit is verified. The obtained data for the non-adiabatic drive is in good agreement with the theoretically predicted results and it converges to the adiabatic reference probability in long time durations confirming the adiabatic theorem. The formulated shortcut protocols are implemented by mapping the NOBIE Hamiltonian to an effective Hamiltonian to drive the qubit. The experimental results of NOBIE method performs as expected and gives adiabatic reference probability irrespective of the time duration of the process. The closeness of the driven qubit state to the adiabatic target state is verified by a Hadamard discrimination and obtained supporting results to confirm adiabaticity. The outputs from the quantum computer always show slight deviation from the theoretically predicted results for both non-adiabatic and shortcut drives. The margin of error from the quantum computer is evident from the vertical spread of the measured outcomes and deviation of its mean value from the theoretically predicted results. The error in the outcomes is mainly due to two reasons, one is the consideration of effective Hamiltonian to describe the system and the other is the effect of possible noise in control pulses. Although the experimental results are prone to error, the outcomes are reliable to compare and confirm the non-adiabatic and shortcut dynamics. This study can be extended to quantum state engineering through an entirely adiabatic path using NOBIE method. Thus, any quantum state can be prepared and manipulated for quantum computation using the NOBIE method in short time duration. Also, developing shortcut protocols for the actual Hamiltonian (with all the interaction terms from other qubits) of the quantum computer will be helpful for more accurate quantum computation.  

{\it Data will be made available on reasonable request  }

\section*{Appendix A}
\label{Appendix A}
\subsection*{Mutually dependent solution}
Multiply the second equation in \autoref{MainConditions} with $z$ and the third equation with $y$. Then, subtracting the second from the third gives
\begin{equation}
y\dot{z}-z\dot{y}=2\left(y^{2}+z^{2}\right)a-2x\left(yb+zc\right).
\end{equation}
 Add and subtract $2x^{2}a$ on right hand side of the above equation to obtain,
 \begin{equation}
 \begin{split}
 y\dot{z}-z\dot{y}&=2\left(x^{2}+y^{2}+z^{2}\right)a-2x\left(xa+yb+zc\right)\\
 &=2f^{2}(t)a-2x\left(xa+yb+zc\right).	
 \end{split}
 \end{equation}
 Rearranging the above equation gives the expression for $a$ as
 \begin{equation}
 a=\frac{1}{2f^{2}(t)}\left(2x\left(xa+yb+zc\right)+\left(y\dot{z}-z\dot{y}\right)\right)
 \end{equation}
Similarly, Multiply the first equation in \autoref{MainConditions} with $z$ and the third equation with $x$. Then, subtract the first from the third to obtain,
\begin{equation}
z\dot{x}-x\dot{z}=2\left(x^{2}+z^{2}\right)b-2y\left(xa+zc\right).
\end{equation}
 Rearranging the above equation after adding and subtracting $2y^{2}b$ on the right-hand side gives
 \begin{equation}
 b=\frac{1}{2f^{2}(t)}\left(2y\left(xa+yb+zc\right)+\left(z\dot{x}-x\dot{z}\right)\right)
 \end{equation}
 Finally, to obtain the expression for $c$, the first equation in \autoref{MainConditions} is multiplied by $y$, and the second equation is multiplied by $x$. Then, subtracting the first from the second gives
 \begin{equation}
 x\dot{y}-y\dot{x}=2\left(x^{2}+y^{2}\right)c-2z\left(xa+yb\right).
 \end{equation} 
 Add and subtract $2z^{2}c$ on the right-hand side to get
 \begin{equation}
 x\dot{y}-y\dot{x}=2\left(x^{2}+y^{2}+z^{2}\right)c-2z\left(xa+yb+zc\right).
 \end{equation}
 Rearranging the above equation to obtain
 \begin{equation}
 c=\frac{1}{2f^{2}(t)}\left(2z\left(xa+yb+zc\right)+\left(x\dot{y}-y\dot{x}\right)\right)
 \end{equation}
 \subsection*{Mutually independent solution}
 Multiply the second equation in \autoref{MainConditions} with $b$ and multiply the third with $c$. Then, add both the equations to get
 \begin{equation}
 f(t)\left(b\dot{y}+c\dot{z}\right)-\dot{f}(t)\left(by+cz\right)=2f(t)\left(yc-zb\right)a
 \end{equation}
 Substituting the first equation in \autoref{MainConditions} on the right-hand side gives
 \begin{equation}
 f(t)\left(b\dot{y}+c\dot{z}\right)-\dot{f}(t)\left(by+cz\right)=-\left(f(t)\dot{x}-\dot{f}(t)x\right)a
 \end{equation}
 rearrange the above equation to obtain a general condition,
 \begin{equation}
 a\left(f(t)\dot{x}-\dot{f}(t)x\right)+b\left(f(t)\dot{y}-\dot{f}(t)y\right)+c\left(f(t)\dot{z}-\dot{f}(t)z\right)=0.
 \label{GeneralConditionforMIS}
 \end{equation}
 From the above equation, we can find any one of the parameters $a,b,$ and $c$ in terms of the other two parameters. To reduce the inverse engineered Hamiltonian to $\sigma_{y}$ and $\sigma_{z}$ components, we have to substitute for $c$ from the above general equation to the second equation in \autoref{MainConditions}, which gives
 \begin{equation}
 \frac{\left(f(t)\dot{y}-\dot{f}(t)y\right)\left(f(t)\dot{z}-\dot{f}(t)z\right)}{f(t)}=2\left(xa\left(\dot{f}(t)x-f(t)\dot{x}\right)+xb\left(\dot{f}(t)y-f(t)\dot{y}\right)-za\left(f(t)\dot{z}-\dot{f}(t)z\right)\right)
 \end{equation} 
Substitute for $a$ from the third equation in \autoref{MainConditions} to get
\begin{equation}
\begin{split}
y\frac{\left(f(t)\dot{y}-\dot{f}(t)y\right)\left(f(t)\dot{z}-\dot{f}(t)z\right)}{f(t)}&=2xb\left[x\left(\dot{f}(t)x-f(t)\dot{x}\right)+y\left(\dot{f}(t)y-f(t)\dot{y}\right)
+z\left(\dot{f}(t)z-f(t)\dot{z}\right)\right]\\
&+\frac{\left(f(t)\dot{z}-\dot{f}(t)z\right)}{f(t)}\left[x\left(\dot{f}(t)x-f(t)\dot{x}\right)+z\left(\dot{f}(t)z-f(t)\dot{z}\right)\right]
\end{split}
\end{equation}
Taking the second term on the right-hand side to the left-hand side of the equation results
\begin{equation}
\begin{split}
-2xb&=\frac{f(t)\dot{z}-\dot{f}(t)z}{f(t)}\\
b&=\frac{\dot{f}(t)z-f(t)\dot{z}}{2xf(t)}
\end{split}
\end{equation}
Substitute the above $b$ to the first equation of \autoref{MainConditions} and subsequent rearrangement using the general condition, \autoref{GeneralConditionforMIS}, gives the expression for $c$ as
\begin{equation}
c=\frac{f(t)\dot{y}-\dot{f}(t)y}{2xf(t)}.
\end{equation}
Finally, substitute the third equation in \autoref{MainConditions} with $b$ to get
\begin{equation}
a=0.
\end{equation}
It is worth noting that we haven't set any of the parameters; $(a,b,c)$, to zero. However, we tried to derive expressions for each of them independent of the other two, which reduced one of the control parameters to achieve STA through the entirely adiabatic path. In the above derivation, we reduced the $\sigma_{x}$ component of control ($a=0$) from the NOBIE Hamiltonian. Selection of a different equation from \autoref{MainConditions} after \autoref{GeneralConditionforMIS} leads to the reduction of any one of the other components of Pauli matrices from $H_{N}$.

\section{Appendix B}
\label{Appendix B}
\subsection*{Calculation of $\Omega_{c}$ using Rabi oscillations}
\begin{figure}[H]
\centering
\includegraphics[width=1\textwidth]{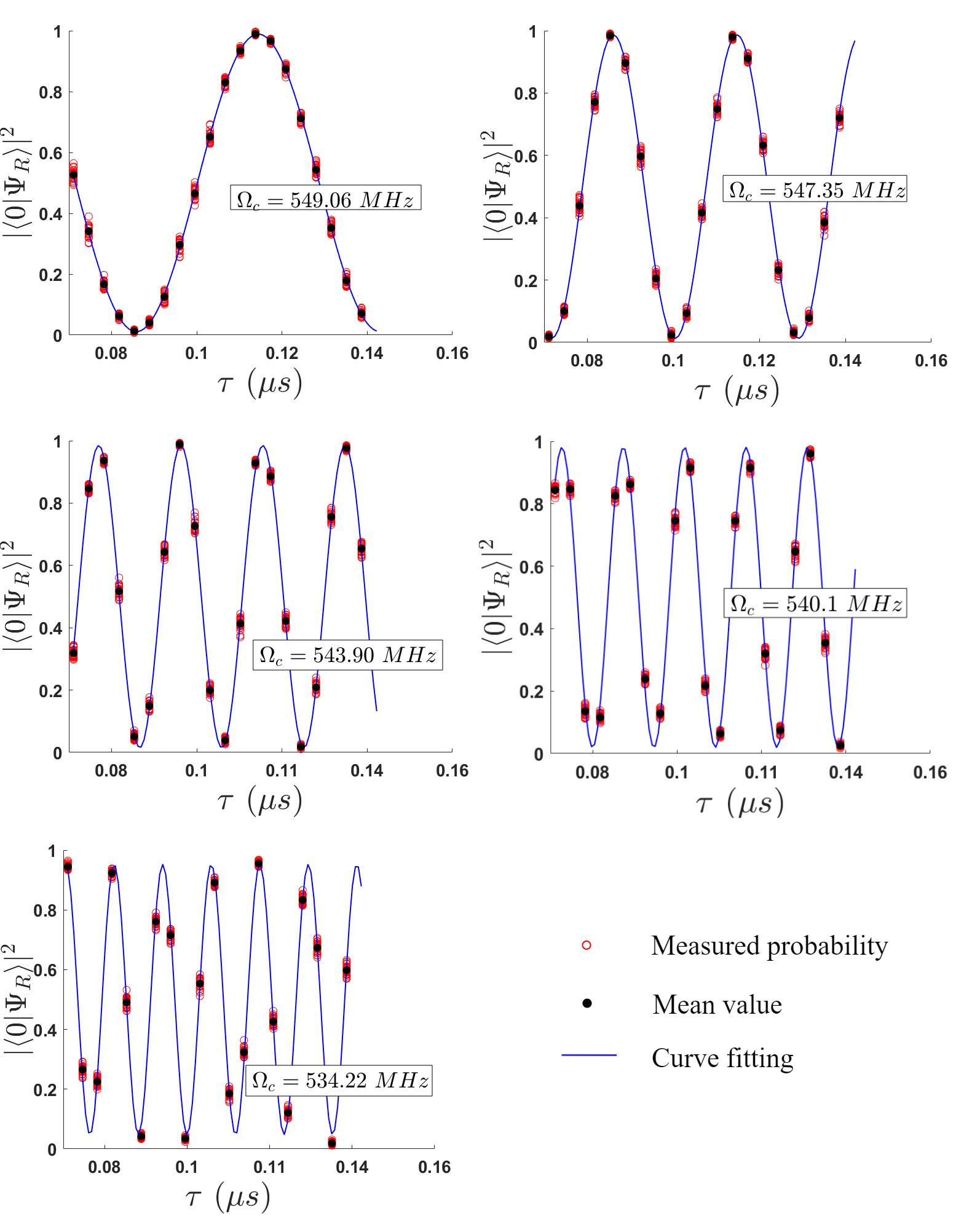}
\caption{}
\label{Rabi_Combine}
\end{figure}

\begin{figure}[H]
\centering
\includegraphics[width=0.75\textwidth]{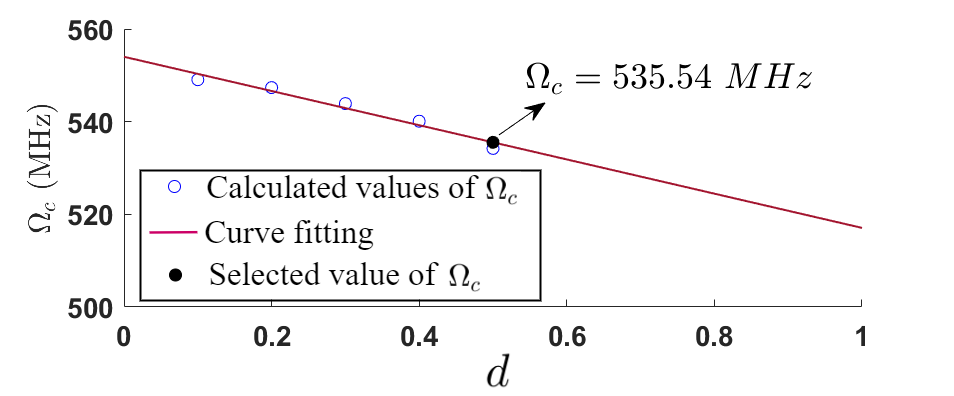}
\caption{}
\label{Rabi_Mean_Oslo}
\end{figure}

The effective Hamiltonian of the system is 
\begin{equation}
H_{eff}=-\frac{1}{2}\Delta \sigma_{z}-d(t)\Omega_{c}\sigma_{x},
\label{EffHamil}
\end{equation}
where $d(t)$ can take values from $0$ to $1$ and $\Omega_{c}$ is the constant amplitude that can be obtained using Rabi oscillations. In this section, we will use constant pulses ($d(t)=d=constant$) to determine $\Omega_{c}$. Then, In a resonant drive of the qubit, i.e., $\Delta=\omega_{q}-\omega_{d}=0$, the qubit state vector will precess about the $x$ axis of the bloch sphere with a frequency of $2d \Omega_{c}$. This precession of the state vector leads to Rabi oscillation among the computational basis states $\vert 0\rangle$ and $\vert 1\rangle$. The schematic diagram of the precession of the state vector is depicted in Figure A1. An expression for $\Omega_{c}$ is obtained by considering the time period ($T$) of the oscillation,
\begin{equation}
\Omega_{c}=\frac{\pi}{Td}.
\label{OmegaC}
\end{equation}
We will execute the resonant drive of the qubit using constant pulses corresponding to different values of $d$ from $0.1$ to $0.5$. Assuming $\vert \Psi_{R}\rangle$ as the evolved state during the above drive, the obtained Rabi oscillations of the probability of $\vert \Psi_{R}\rangle$ to be in the $\vert 0\rangle$ is given in \autoref{Rabi_Combine}. The measured probabilities for any particular time duration is spreded vertically indicating the error in the measured output state. This error primary due to the assumption of an effective Hamiltonian for a single qubit, while the actual qubit Hamiltonian includes the interaction terms with other qubits. Also, there are chances of reset errors and noise in the driving pulse. We select the mean value of the measured probabilities for all the time durations plotted and fit a cosine curve to get the Rabi Oscillations. The time period of the fitted curve can be used to get the value of $\Omega_{c}$ using the \autoref{OmegaC}. The values of $\Omega_{c}$ obtained are given in each of the subfigures corresponding to different $d$ values. It is ideal get a constant value for $\Omega_{c}$ irrespective of $d$ values, however we observed a decreasing treand with increasing $d$ value, i.e., the increase in the amplitude of the constant pulse. This trend is expected as we are using an effective Hamiltonian assuming the qubit is isolated from the other qubits of the quantum computer. The higher values for the pulse amplitude, $0.5\leq d\leq 1$ are avoided to decrease the effect of amplitude noise in the calculation of $\Omega_{c}$. Also, to select an appropriate value for $\Omega_{c}$, we have fitted a straight line to the calculated values of $\Omega_{c}$ (see \autoref{Rabi_Mean_Oslo} use the mean value of the straight line function of the domain, $d\in [0,1]$. The selected mean value for $\Omega_{c}$ is indicated as a black dot in \autoref{Rabi_Mean_Oslo} ($\Omega_{c}=535.54~MHz$). The error in the measured output states (the vertical spread of probabities), the decreasing trend of the calculated $\Omega_{c}$, and the error in the selection of $\Omega_{c}$ might lead to errors in any arbitrary evolution of the qubit using the Hamiltonian in \autoref{EffHamil}. Expecting such error in both non-adiabatic and NOBIE methods, the selected value of $\Omega_{c}$ is enough to qualitatively analyse the non-adiabatic and shortcut evolution of the qubit. 
\begin{figure}[H]
\centering
\includegraphics[width=0.5\textwidth]{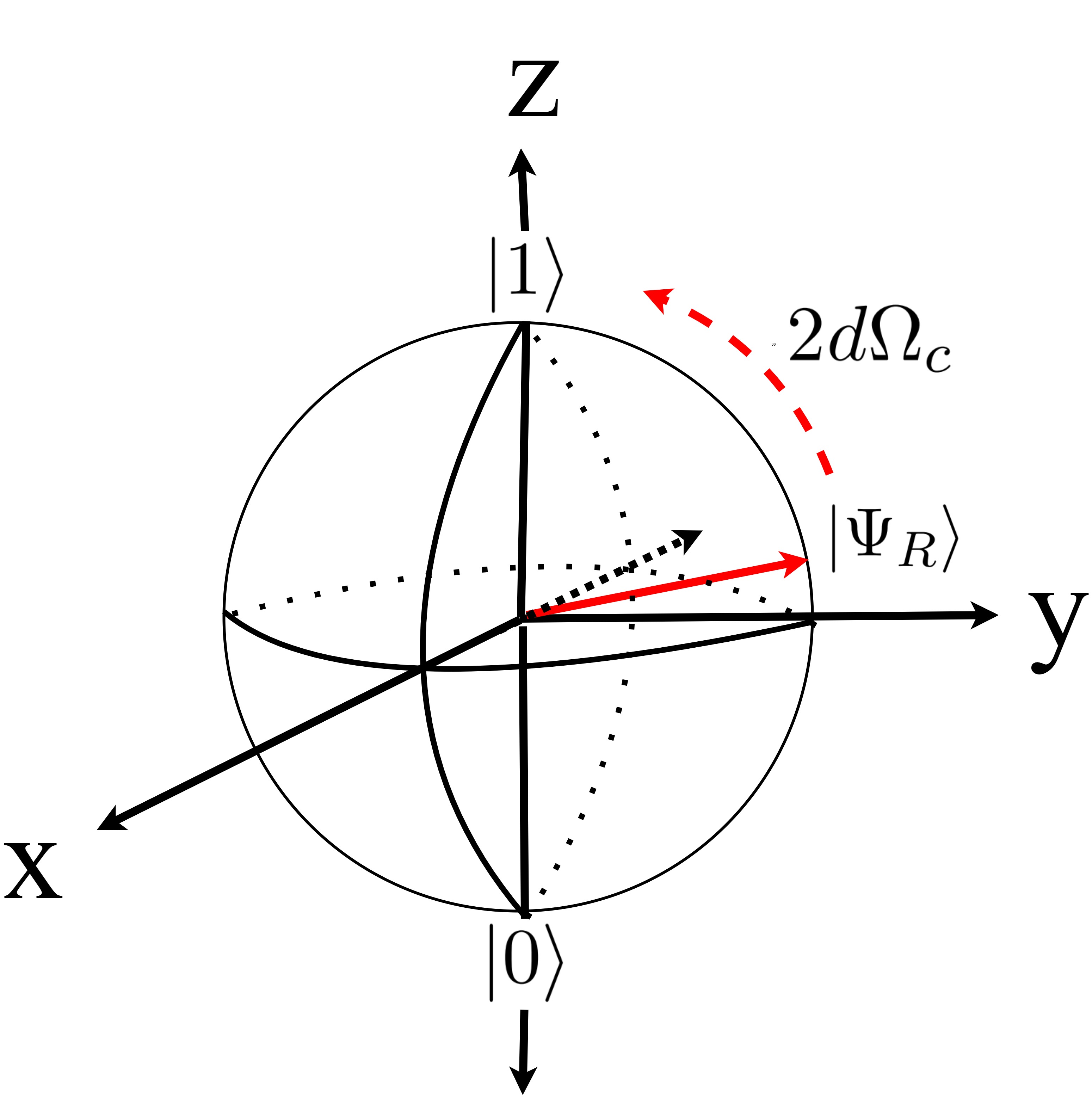}
\caption{}
\label{Blochsphere}
\end{figure}

\end{document}